\documentclass[a4paper,3p,times]{elsarticle}
\usepackage[utf8]{inputenc}
\usepackage{amsmath}
\usepackage{amsfonts}
\usepackage{graphicx}
\usepackage{float}
\usepackage{bm}
\usepackage{hyperref}
\usepackage{color}
\usepackage{amssymb}
\usepackage{dsfont}
\usepackage{slashed}
\usepackage{mathtools}
\usepackage{hyperref}
\hypersetup{
colorlinks=true,
urlcolor=blue,
linkcolor = blue,
urlcolor  = blue,
citecolor = blue,
anchorcolor = blue,
pdftitle={triton lifetime},
pdfsubject={triton lifetime}}
\allowdisplaybreaks
\makeatletter
\def\ps@pprintTitle{%
\let\@oddhead\@empty
\let\@evenhead\@empty
\def\@oddfoot{}%
\let\@evenfoot\@oddfoot}
\makeatother
\begin{document}
\title{The triton lifetime from nuclear lattice effective field theory}
%%%%%%%%%%%%%%%%%%%%%%%%%%%%%%%%%%%%%%%%%%%%%%%%%%%%%%
\author[gazin,bonn]{Serdar Elhatisari}
\author[fzj]{Fabian Hildenbrand}
\author[bonn,fzj,tbilisi]{Ulf-G. Mei\ss{}ner}
%%%%%%%%%%%%%%%%%%%%%%%%%%%%%%%%%%%%%%%%%%%%%%%%%%%%%%
\address[gazin]{Faculty of Natural Sciences and Engineering, Gaziantep Islam
Science and Technology University, Gaziantep 27010, Turkey}
\address[bonn]{Helmholtz-Institut f\"ur Strahlen- und
Kernphysik and Bethe Center for Theoretical Physics,\\
Universit\"at Bonn, D-53115 Bonn, Germany}
\address[fzj]{Institute for Advanced Simulation (IAS-4), Forschungszentrum J\"ulich,
D-52425 J\"ulich, Germany}
\address[tbilisi]{Tbilisi State University, 0186 Tbilisi, Georgia}

%%%%%%%%%%%%%%%%%%%
%%%%%%%%%%%%%%%%%%%
\begin{abstract}
In this work, we present a  calculation of the triton $\beta$-decay lifetime using Nuclear Lattice Effective Field Theory (NLEFT) at next-to-next-to-next-to-leading order in the chiral expansion. By incorporating a non-perturbative treatment
of the higher-order corrections, we achieve consistent predictions for the Fermi and Gamow-Teller matrix elements, which are crucial for determining the triton lifetime. Our results are consistent with earlier theoretical calculations, confirming the robustness of our approach. This study marks a significant advancement in the systematic application of NLEFT to nuclear $\beta$-decay processes, paving the way for future high-precision calculations in more complex nuclear systems. Additionally, we discuss potential improvements to our approach, including the explicit inclusion of two-pion exchange mechanisms and the refinement of three-nucleon forces. These developments are essential for extending the applicability of NLEFT to a broader range of nuclear phenomena, including neutrinoless double-$\beta$ decay.
\end{abstract}
%%%%%%%%%%%%%%%%%%%
%%%%%%%%%%%%%%%%%%%
\maketitle
%%%%%%%%%%%%%%%%%%%%%%%%%%%%%%%%%%%%%%%%%%%%%%%%%%%%%%

\section{Introduction}
\label{sec:intro}

\begin{figure}[h]
\centering
\includegraphics[width=0.47\linewidth]{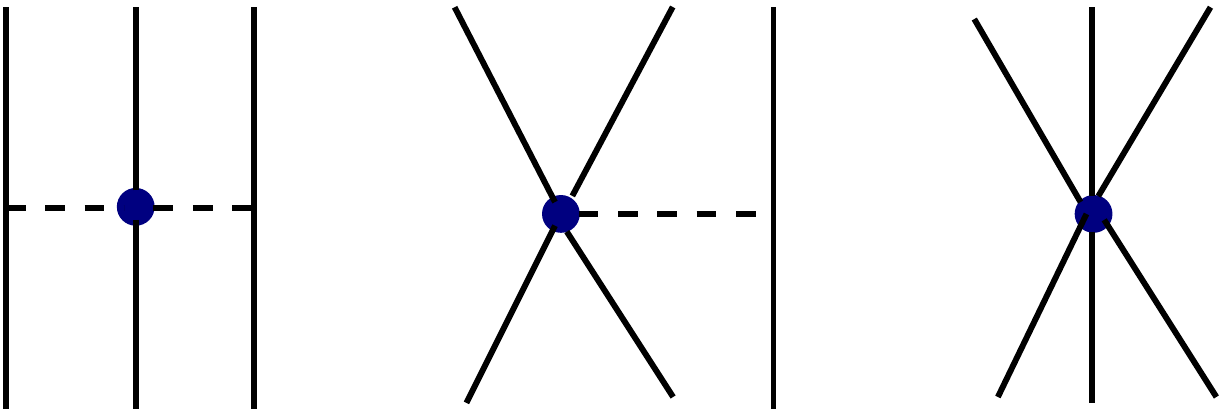}
\caption{Topologies of the leading order 3NFs in chiral EFT: Two-pion exchange
(left), one-pion exchange (middle) and six-fermion contact term (right).
\label{fig:topo}}
\end{figure}

Three-nucleon forces (3NFs) play an important role in precision calculations of
nuclei and nuclear matter, for reviews see
e.g. Refs.~\cite{Epelbaum:2008ga,Kalantar-Nayestanaki:2011rzs,Hammer:2012id,Epelbaum:2019kcf}.
Within chiral effective field theory (EFT) as advocated by Weinberg, 3NFs appear at
next-to-next-to-leading order (N2LO) \cite{Weinberg:1991um,vanKolck:1994yi,Friar:1998zt} in terms
of the three topologies displayed in Fig.~\ref{fig:topo}. These different terms come
with low-energy constants (LECs).
In case of the two-pion exchange, these LECs that are called $c_{1,3,4}$. They can be precisely
determined from pion-nucleon scattering~\cite{Hoferichter:2015tha}, demonstrating the
power of chiral symmetry in connecting seemingly unrelated processes. To pin down the
LECs $c_D$ and $c_E$, that parameterize the one-pion exchange and contact term topologies
at this order, respectively, one must consider observables in three-nucleon systems.
In a first systematic analysis of neutron-deuteron scattering at N2LO, these LECs
were fixed from the triton binding energy and the neutron-doublet scattering length
\cite{Epelbaum:2002vt}, although these quantities display some correlation.
It was pointed out first in Ref.~\cite{Gazit:2008ma} that triton $\beta$-decay, that
is the triton lifetime, together with the binding energies in the $A=3$ system
can lead to a robust determination of $c_D$ and $c_E$. It could later be shown
that using the cross section minimum in low-energy proton-deuteron as well as
neutron-deuteron scattering
can also leads to a fairly precise determination of the short-distance LEC $c_D$,
see Ref.~\cite{LENPIC:2018ewt} (and references therein). It is worth pointing out that
in these continuum approaches, a completely consistent regularization scheme for
two- and three-body forces as well as external currents has only become available
very recently~\cite{Krebs:2023gge}. In pionless EFT, where all interactions
are represented by contact terms, triton $\beta$-decay has been studied in Ref.~\cite{De-Leon:2016wyu}.

Here, we approach the problem of triton $\beta$-decay from a different perspective,
namely in the framework of Nuclear Lattice Effective Field Theory (NLEFT). For an introduction to that method, see
Refs.~\cite{Lee:2008fa,Lahde:2019npb}. This lattice approach has proven successful
in solving problems that were previously considered intractable with conventional methods,
like the first {\em ab initio} calculations of the Hoyle state in the
spectrum of $^{12}$C~\cite{Epelbaum:2011md} and of
alpha-alpha scattering~\cite{Elhatisari:2015iga}. These calculations were
performed at N2LO on a coarse lattice, which limits the theoretical precision. A major step forward in achieving high precision in NLEFT for many-nucleon systems was recently made using the so-called the wavefunction matching method at next-to-next-to-next-to-leading order
(N3LO) in the chiral expansion~\cite{Elhatisari:2022zrb}. High-fidelity chiral interactions at N3LO often encounter significant sign problems due to the cancellation of positive and negative contributions, making Monte Carlo calculations impractical. The wavefunction matching method introduced in Ref.~\cite{Elhatisari:2022zrb} resolves this issue for two-nucleon interactions at N3LO. This method was successfully applied to light nuclei,
medium-mass nuclei ($A \leq 58$), neutron matter, and nuclear matter, and good
agreement with the empirical data was found. Despite the success of the wavefunction matching method in improving theoretical precision, the calculations in Ref.~\cite{Elhatisari:2022zrb} were carried out using first-order perturbation theory. Since first-order perturbation theory only provides corrections to the energy and not to the wavefunctions, triton $\beta$-decay calculations at N3LO, requiring higher-order perturbative corrections for realistic wave functions, cannot be directly performed using the methods from Ref.~\cite{Elhatisari:2022zrb}.  One potential solution to this challenge is to extend the calculations to second-order perturbation theory. Recent advances in perturbative quantum Monte Carlo (QMC) methods, as detailed in Ref.~\cite{Lu:2021tab}, provide an effective framework for incorporating higher-order perturbative corrections, making it particularly well-suited for applications to heavier nuclei. Alternatively, fully non-perturbative methods can be applied to light nuclear systems to generate realistic wave functions at N3LO, as required for triton $\beta$-decay calculations. In this paper, we adopt this non-perturbative approach.

In our calculations, we use the same action as in Ref.~\cite{Elhatisari:2022zrb}. Our findings indicate that for a given set of LECs, our theoretical precision aligns well with earlier theoretical calculations in continuum, allowing us to explore variations of the smeared $c_D$ and $c_E$ LECs and their impact on the triton lifetime. It should also be stressed that the triton $\beta$-decay offers a benchmark for investigating
the  weak interactions in nuclei, eventually paving the way for the calculation of the
neutrinoless double-$\beta$ decay in nuclei like e.g. $^{48}$Ca or $^{76}$Ge. By incorporating the wavefunction matching method and advances in perturbative quantum Monte Carlo, NLEFT is now equipped to address these more complex systems, paving the way for high-precision studies of heavier nuclei.

This article is organized as follows. In Sec.~\ref{sec:form} we present the
underlying formalism and the treatment of the 3NFs. In Sec.~\ref{sec:res},
we display and discuss the results on the various matrix elements in triton
$\beta$-decay and their sensitivity to the LECs.  We end with a short summary
and outlook in Sec.~\ref{sec:summ}. In~\ref{app}, we display the two- and three-nucleon forces on the
lattice as they are used here.

%%%%%%%%%%%%%%%%%%%%%%%%%%%%%%%%%%%%%%%%%%%%%%%%%%%%%%
\section{Formalism}
\label{sec:form}
%%%%%%%%%%%%%%%%%%%%%%%%%%%%%%%%%%%%%%%%%%%%%%%%%%%%%%

Triton $\beta$-decay is the process where $^3$H decays into $^3$He, an electron, and an electron antineutrino,
\begin{equation}
^3{\rm H} \to {}^3{\rm He} + e^- + \bar{\nu}_e~.
\end{equation}
The matrix elements of the weak transition are crucial to understanding this decay process. This section outlines the formalism used in calculating the relevant matrix elements, including the Fermi ${F} \sim \tau$ and the Gamow-Teller ${GT} \sim \sigma\tau$ operators, which are essential for describing the nuclear structure and weak interaction dynamics involved in triton  $\beta$-decay. Here, $\tau$ and $\sigma$ are the nucleon isospin and spin operators, respectively.

The half-life of triton $\beta$-decay, $t_{1/2}$, can be expressed in terms of the Fermi and Gamow-Teller matrix elements as follows~\cite{Schiavilla:1998je},
\begin{equation}
(1+\delta_{R}) \, t_{1/2} \, f_{V} = \frac{K/G_{V}^{2}}{\langle{F}\rangle^2
+ ({f_{A}}/{f_{V}}) \, g_{A}^{2} \,\langle{GT}\rangle^2}\,,
\end{equation}
where $K = 2\pi^3 \ln 2 / m_e^5$ (with $m_e$ the electron mass), $G_V$ is the weak interaction vector coupling constant, $f_V = 2.8355\times 10^{-6}$ and $f_A = 2.8506\times 10^{-6}$ are the Fermi functions~\cite{Simpson:1987zz}, $g_A=1.287$ is the axial coupling constant, and $\delta_R$ accounts for radiative corrections of $1.9\%$~\cite{Raman:1978qta}. The comparative half-life of triton, $(1+\delta_{R}) \, t_{1/2} \, f_{V}$, is taken as $1129.6(30)$~\text{s}~\cite{Akulov:2005umb} ($1134.6(30)$~\text{s}~\cite{Baroni:2016xll}), and the value of $K/G_V^2$ is $6146.6(6)$~s~\cite{Hardy:1990sz} ($6144.5(19)$~s~\cite{Hardy:2014qxa}).

The matrix elements $\langle F \rangle$ and $\langle GT \rangle$ are calculated using the wave functions of $^3$H and $^3$He. These wave functions can be obtained employing the hyperspherical-harmonics expansion method~\cite{Baroni:2016xll} or the
non-perturbative Faddeev equations~\cite{De-Leon:2019dqq} from high-precision nuclear interactions, such as the Argonne v18 two-nucleon potential and the Urbana-IX three-nucleon potential as well as chiral interactions. For the Fermi operator, the calculation based on the Argonne v18 two-nucleon
forces supplemented with the Urbana-IX 3NF gives
\begin{equation}
\label{eq:Fold}
\langle{F}\rangle
= \sum_{n = 1}^{3} \langle{}^3{\rm He}\|\tau_{n,+}\|{}^3{\rm H}\rangle = 0.9998~,
\end{equation}
where $\tau_{n,+}$ is the isospin-raising operator. This value indicates a near-perfect overlap of the isospin components between the initial and final states, with minor corrections due to charge-symmetry breaking and electromagnetic effects in the nuclear interaction~\cite{Baroni:2016xll}. The Gamow-Teller matrix element, which involves both spin and isospin operators, is expressed as,
\begin{equation}
\langle GT \rangle = \sum_{n=1}^3 \langle ^3\text{He} || \sigma_n \tau_{n,+} || ^3\text{H} \rangle~.
\end{equation}
The empirical value of the Gamow-Teller operator can be deduced from the triton lifetime,
\begin{equation}
\langle GT \rangle_{\text{emp}} = \sqrt{  \frac{(K/G_V^2)/[(1 + \delta_R)\, t_{1/2} \,f_V] - \langle F \rangle^2}{({f_A}/{f_V})\, g_A^2}} = 1.6497(23) \,.
\end{equation}
This value is obtained by fitting the theoretical predictions to the experimental data, including corrections for meson-exchange currents (MECs) and relativistic effects. We use here and for the remainder of the paper the averaged values of Refs.~\cite{Akulov:2005umb,Baroni:2016xll} for $(1+\delta_{R}) \, t_{1/2} \, f_{V} = 1132.1 (25)$~and from Refs.~\cite{Hardy:1990sz,Hardy:2014qxa} for $K/G_V^2 = 6145.5 (11)$\,, respectively. It should be noted that in this determination of the Gamov-Teller matrix element, theoretical
input is used to pin down the Fermi matrix element. Here, we will both calculate $\langle{F}\rangle$ and $\langle{GT}\rangle$ in a consistent scheme, namely NLEFT, and then predict the triton lifetime.

Our approach to triton $\beta$-decay calculations employs NLEFT, and we briefly discuss the main ingredient of the method used here and refer to
Ref.~\cite{Elhatisari:2022zrb} for more details. In a nutshell, a new
quantum many-body approach, the so-called wave function matching, transforms the high-fidelity interaction between particles
so that the wave functions of the high-fidelity Hamiltonian up to some finite range match that of an easily
computable Hamiltonian. More precisely, wavefunction matching operates entirely in the two-nucleon
sector. For the nuclear case, this simplified Hamiltonian
consists of Wigner SU(4) symmetric two-nucleon forces and
the properly regularized one-pion exchange, and it is treated fully non-pertubatively. To bring the chiral Hamiltonian $H$ close to the simplified Hamiltonian $H_S$,
a unitary transformation is performed leading to $H' = U^\dagger HU$, and the differences to the full chiral Hamiltonian,  $H'-H_S$, are then calculated in first order perturbation theory. Finally, fitting the various locally and non-locally smeared 3NF operators to the nuclear binding energies with $3\le A \le 58$, one can predict the corresponding nuclear charge radii as well as the equation of state of pure neutron as well as nuclear matter. All of these quantities agree with the data. In this work, we extend beyond the first-order perturbation theory previously used for calculations at N3LO in the chiral effective field theory~\cite{Elhatisari:2022zrb}, applying fully non-perturbative calculations to three-nucleon systems, with the corresponding Hamiltonian given in~\ref{app}.
This approach allows us to obtain realistic wave functions necessary for accurate triton $\beta$-decay calculations at N3LO. Our method involves solving the Schrödinger equation for the three-nucleon system using the Lanczos eigenvector method~\cite{Lanczos:1950zz} , which has been successfully applied in nuclear lattice calculations. For a detailed discussion of the method, we refer the reader to Ref.~\cite{Stellin:2018fkj}, and for its early application in NLEFT, see Ref.~\cite{Borasoy:2005yc}. The three-dimensional space is represented by a finite volume $L\times L\times L$, with $L$ the spatial extension, and  with lattice spacing $a=1.32$~fm in the spatial direction. This corresponds to the magic
momentum cutoff of $p_{\rm max}\simeq 465\,$MeV that best displays the hidden
spin-isospin symmetry of QCD~\cite{Lee:2020esp}. %To achieve the required
%precision, fully non-perturbative methods are applied. %%%%%%%%%%%%%%%%%%%%%%%%%%%%%%%%%%%%%%%%%%%%%%%%%%%%%%%%%%%%%%%%%%%%%%
\section{Results and discussion}
\label{sec:res}
%%%%%%%%%%%%%%%%%%%%%%%%%%%%%%%%%%%%%%%%%%%%%%%%%%%%%%%%%%%%%%%%%%%%%%

This section presents the results of our calculation of the triton lifetime using NLEFT. We focus on the determination of the Fermi and Gamow-Teller matrix elements, $\langle F \rangle$ and $\langle GT \rangle$, and analyze their dependence on the strength of the 3NFs. These results are compared with experimental data and previous theoretical studies to evaluate the accuracy and reliability of our approach.

\begin{figure}[htb]
\centering
\includegraphics[width=0.45\linewidth]{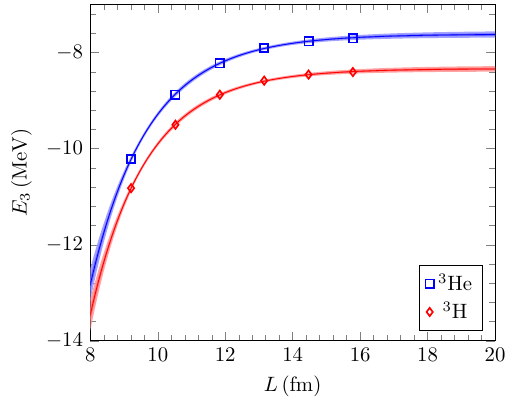}~~~
\includegraphics[width=0.45\linewidth]{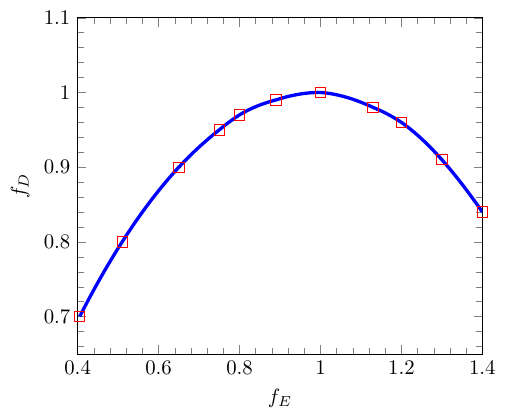}
\caption{The ground state energies of $^3$He and $^3$H nuclei as a function of the box size $L$ (left panel). The trajectory of $f_D$ -- $f_E$ that leaves these energies invariant (right panel). The dark (light)
red/blue bands refer to the 90~(99)~\% confidence level of the extrapolation. In the right panel the red squares display the result of our calculation and the blue line is a 3$^{\rm rd}$ order spline interpolation to guide the eye.
\label{fig:en.tra}}
\end{figure}

As noted before, we employ the same lattice action as in Ref.~\cite{Elhatisari:2022zrb}, with the key difference being the non-perturbative treatment of the interactions. Our calculations start with the determination of the ground state energies of ${}^3$H and ${}^3$He. Using the NLEFT framework, we obtain  infinite-volume extrapolated ground state energies of $8.33(2)$~MeV and $7.62(2)$~MeV, respectively, as shown in the left panel of Fig.~\ref{fig:en.tra}. These values are slightly below the empirical values of $8.48$~MeV and $7.72$~MeV, respectively. %Since the energies %%at $L=12$ (15.84~fm) are very close to the extrapolated results, %we present the results at $L=12$ for all further analyses.

Next, we calculate the Fermi and Gamow-Teller matrix elements, $\langle F \rangle$ and $\langle GT \rangle$, and examine their sensitivity to the strengths of the three-nucleon one-pion exchange and three-nucleon SU(4) symmetric potentials. The representation of these two 3NF topologies on the lattice differs from that in the continuum, as the corresponding 3NFs take the form~\cite{Elhatisari:2022zrb},
\begin{align}
V_{c_{D}}
= f_{D} \left( V_{c_{D}}^{(0)} +
  V_{c_{D}}^{(1)} +
  V_{c_{D}}^{(2)}
   \right)
\,,
\label{eqn:V_cD--001}
\end{align}
\begin{align}
V_{c_{E}}
= f_{E} \left( V_{c_{E}}^{(0)} +
  V_{c_{E}}^{(1)} +
  V_{c_{E}}^{(2)} +   V_{c_{E}}^{(l)} +
  V_{c_{E}}^{(t)}
   \right)
\,,
\label{eqn:V_cE--001}
\end{align}
where the explicit forms of $V_{c_D}^{(n)}$ and $V_{c_E}^{(n)}$ are given in~\ref{app}. It is important to note that all these potential terms, with their various forms of smearing, are associated with the LECs $c_D^{(n)}$ and $c_E^{(n)}$, also listed in~\ref{app}, making a direct comparison with the continuum values of these LECs impossible. Here, in Eqs.~(\ref{eqn:V_cD--001}) and (\ref{eqn:V_cE--001}), we introduce the factors $f_D$ and $f_E$ to allow us to vary the overall contributions from the one-pion-exchange potential $V_{c_D}$ and the SU(4) symmetric short-range potential $V_{c_E}$. We then explore the trajectory of $f_D$ versus $f_E$ that maintains the binding energies of the three-body systems invariant~\cite{Epelbaum:2002vt,Gazit:2008ma}.
In the following analysis, we vary $f_E$ within the range $[0.4,1.4]$, which, due to the aforementioned correlation, causes $f_D$ to vary from $0.7$ to $1.0$, as shown in the right panel of Fig.~\ref{fig:en.tra}. Because the potential terms in Eqs.~(\ref{eqn:V_cD--001}) and (\ref{eqn:V_cE--001}) respond differently to the three-body systems, the trajectory observed in Fig.~\ref{fig:en.tra} is more complex than the one presented in Ref.~\cite{Gazit:2008ma}.

\begin{figure}[htb]
\centering
\includegraphics[width=0.46\linewidth]{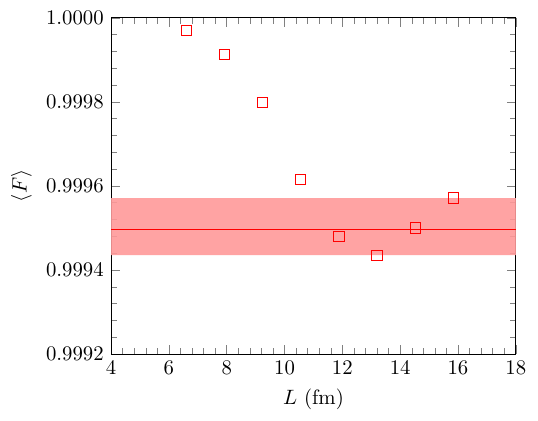}
\includegraphics[width=0.45\linewidth]{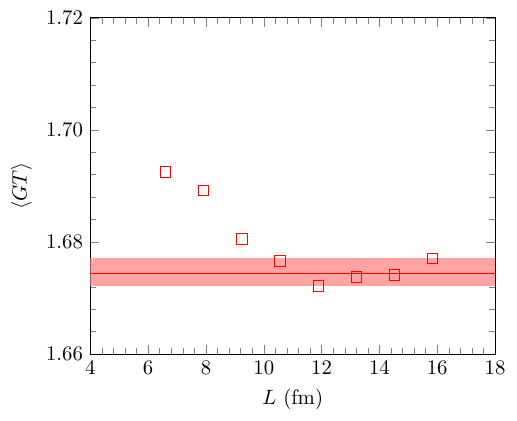}
\caption{Plots of Fermi (left panel) and Gamow-Teller (right panel) matrix elements  at a function of $L$ and the extraction of the corresponding plateaus.
\label{fig:plateau}}
\end{figure}

\begin{figure}[htb]
\centering
\includegraphics[width=0.32\linewidth]{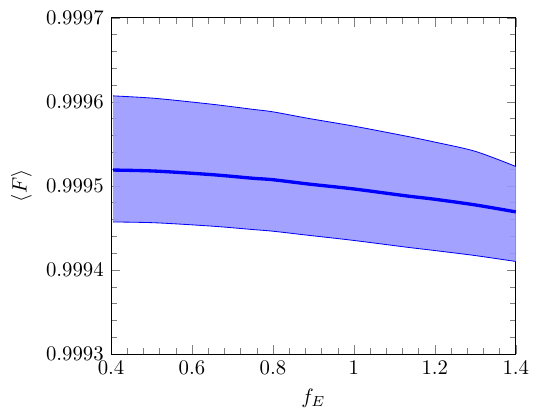}
\includegraphics[width=0.32\linewidth]{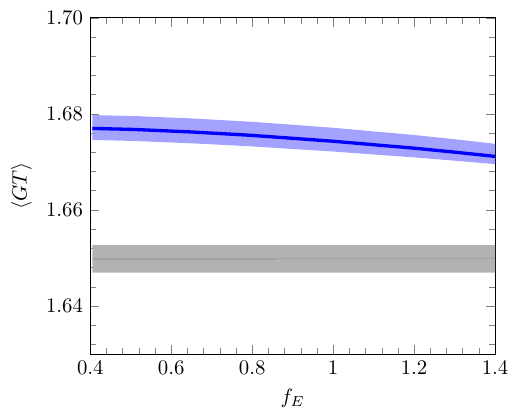}
\includegraphics[width=0.32\linewidth]{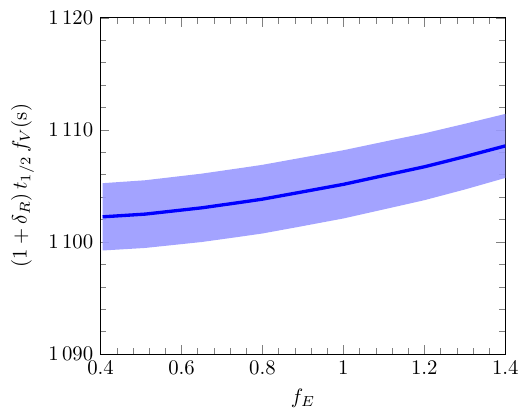}
\caption{Plots of Fermi (left panel) and Gamow-Teller (middle panel) matrix elements and the triton lifetime (right panel) as a function of the parameter $f_E$ as defined in Eq.~\eqref{eqn:V_cE--001}.  The blue bands represents uncertainties estimated by a time-honoured plateau. The grey band in the middle panel refers to using the averaged empirical values together with the calculated value of $\langle F\rangle$ (left panel). The impact of the uncertainties of $\langle F\rangle$ are marginal as indicated by the black band inside the gray band.
\label{fig:f.Gt}}
\end{figure}

Since no established infinite-volume extrapolation methods exist for matrix elements involving different initial and final states, we employ the well-known plateau method from lattice QCD to extract the values of $\langle F \rangle$ and $\langle GT \rangle$. This approach is illustrated in Fig.~\ref{fig:plateau}. Thus, the Fermi operator as well as the Gamow-Teller operator are simultaneously obtained within the consistent framework of NLEFT,  and we obtain,
 %final results at $L=12$ are,
\begin{eqnarray}
\langle F\rangle &=& 0.99949(11)\, , \label{eq:F}\\
\langle GT\rangle &=& 1.6743(58)\, \label{eq:GT} .
\end{eqnarray}
Furthermore, in Fig.~\ref{fig:f.Gt}, these matrix elements are shown  as a function of $f_E$.

The uncertainty estimate is based on the variation of the operators over the range of $f_E$ and $f_D$, as shown in Fig.~\ref{fig:f.Gt}, encompassing sets of three-body forces that describe the three-body systems equally well together with the uncertainties induced by the widths of the plateaus in Fig.~\ref{fig:plateau}. A similar dependence on the three-body forces was observed in Ref.~\cite{Gazit:2008ma}. Notably, our value of $\langle F \rangle$ is close to the one given in Eq.~\eqref{eq:Fold}.
The corresponding lifetime is given by
\begin{equation}
\label{eq:lt}
(1+\delta_{R}) \, t_{1/2} \, f_{V} =1105.1(74)~\text{s},
\end{equation}
where the uncertainties are also inherited from $K/G^2_V$, but dominated by the uncertainties for $\langle GT\rangle$. In the right panel of Fig.~\ref{fig:f.Gt}, we show the dependence of the lifetime on the coupling $f_E$. Additionally, we calculate $\langle GT_F \rangle$ using $\langle F \rangle$ from Eq.~\eqref{eq:F} and the averaged empirical half-lives from Refs.~\cite{Akulov:2005umb,Baroni:2016xll}, injected into Eq.~\eqref{sec:form}, giving (see also the middle panel in Fig.~\ref{fig:f.Gt}),
\begin{equation}
\langle GT_F\rangle = 1.6498(29)\,.
\end{equation}

%%%%%%%%%%%%

%%%%%%%%%%%%%%%%%%%%%%%%%%%%%%%%%%%%%%%%%%%%%%%%%%%%%%%%%%%%%%%%%%%%%%
\section{Summary and outlook}
\label{sec:summ}
%%%%%%%%%%%%%%%%%%%%%%%%%%%%%%%%%%%%%%%%%%%%%%%%%%%%%%%%%%%%%%%%%%%%%%

In this study, we have performed a detailed calculation of triton $\beta$-decay using NLEFT with the N3LO Hamiltonian as developed in Ref.~\cite{Elhatisari:2022zrb}, incorporating a non-perturbative treatment of the higher-order corrections. With all parameters of the two- and three-nucleon forces already determined, we were able to consistently predict the Fermi and Gamow-Teller matrix elements given in Eqs.~(\ref{eq:F},~\ref{eq:GT}) and the triton lifetime given in Eq.~(\ref{eq:lt}). This work represents a significant step forward in the systematic study of nuclear $\beta$ and double-$\beta$ decays within the framework of NLEFT.

Clearly, there are several avenues for further improvement and exploration to enhance the accuracy and scope of these calculations. First, the explicit inclusion of the two-pion exchange in the two-nucleon interaction, as discussed in Ref.~\cite{Elhatisari:2022zrb}, would enable a more consistent treatment of higher-order corrections to the relevant exchange currents~\cite{Menendez:2011qq}.
Incorporating this mechanism would align our approach more closely with the underlying chiral dynamics of nuclear forces.

Second, further refinement of the three-nucleon forces may be necessary to address any remaining discrepancies that were not apparent in previous studies~\cite{Elhatisari:2022zrb}. This includes exploring the effects of different regularization schemes, which could provide deeper insights into the structure of the 3NFs.

Lastly, the formalism for infinite volume extrapolation of matrix elements, which was not fully addressed in this work, should be developed to reduce finite volume effects and improve the reliability of the lattice calculations. Such advancements would be crucial for extending our approach to more complex nuclear systems, including the study of neutrinoless double-$\beta$ decay in heavier nuclei like $^{48}$Ca and $^{76}$Ge.

In summary, this work represents a significant step forward in the application of NLEFT to nuclear weak decay processes. The methodologies developed and results obtained here lay the groundwork for future studies aimed at achieving high-precision predictions for a wide range of nuclear phenomena. As we continue to refine our theoretical tools and expand the range of systems studied, NLEFT promises to play a crucial role in advancing our understanding of such fundamental nuclear processes.

%%%%%%%%%%%%%%%%%%%%%%%%%%%%%%%%%%%%%%%%%%%%%%%
\section*{Acknowledgements}
SE thanks Lukas Bovermann for useful discussions. This work was supported in part by the European
Research Council (ERC) under the European Union's Horizon 2020 research
and innovation programme (grant agreement No. 101018170).
The work of UGM was also supported in part by the CAS President's International
Fellowship Initiative (PIFI) (Grant No.~2025PD0022) and  by the MKW NRW
under the funding code NW21-024-A.
The work of SE was further supported by
the Scientific and Technological Research Council of Turkey (TUBITAK project no. 116F400 and 120F341).
The EXOTIC project is supported by the Jülich Supercomputing Centre by
dedicated HPC time provided on the JURECA DC GPU partition.
The authors gratefully acknowledge the Gauss Centre for Supercomputing e.V.
(www.gauss-centre.eu) for funding this project by providing computing time on the GCS Supercomputer JUWELS at J\"ulich Supercomputing Centre (JSC).

%%%%%%%%%%%%%%%%%%%%%%%%%%%%%%%%%%%%%%%%%%%%%%%
\begin{appendix}

     \section{Hamiltonian at N3LO}
\label{app}

This section provides some details of our realistic Hamiltonian utilized in the calculations. We have developed the 2NFs in the framework of chiral effective field theory at N3LO, with 24 LECs accurately fitted to match empirical partial wave phase shifts and mixing angles~\cite{Li:2018ymw}.
The 3NFs have been recently included into the framework, by constraining the LECs to some selected nuclear binding energies~\cite{Elhatisari:2022zrb}.

In Ref.~\cite{Li:2018ymw} the 2NFs were constructed using a non-local smearing parameter $s_{\rm NL}$, while in Ref.~\cite{Elhatisari:2022zrb} we have constructed 2NFs using another set of non-local contact operators by introducing a non-local regulator $f_{\Lambda} = \exp[-\sum_{i=1}^{2}(p_{i}^2 + {p_{i}^{\prime}}^2)/\Lambda^2]$, where $p_{i}$ and $p_{i}^{\prime}$ are the momenta of individual nucleons. We define the Hamiltonian $H$ as,
\begin{equation}
H = K + V_{\rm OPE} + V_{\rm Coulomb}
+ V_{\rm 3N}^{\rm Q^3}
+ V_{\rm 2N}^{\rm Q^4}
+ W_{\rm 2N}^{\rm Q^4}\,,
\label{eq:H-N3LO}
\end{equation}
where $K$ is the kinetic energy term constructed using fast Fourier transforms to produce the exact dispersion relation
$E_N = p^2/(2m_{N})$, with the nucleon mass $m_{N}=938.92$~MeV. Further, $V_{\rm OPE}$ is the one-pion-exchange potential defined using the regularization method given in Ref.~\cite{Reinert:2017usi},
\begin{align}
V_{\rm OPE}  = - &   \frac{g_A^2}{8f^2_{\pi}}\ \, \sum_{{\bf n',n},S',S,I}
:\rho_{S',I\rm }^{(0)}(\vec{n}')f_{S',S}(\vec{n}'-\vec{n})   \rho_{S,I}^{(0)}(\vec{n}):   -C_{\pi} \, \frac{g_A^2}{8f^2_{\pi}} \sum_{{\bf n',n},S,I}
:\rho_{S,I}^{(0)}(\vec{n}')
f^{\pi}(\vec{n}'-\vec{n})
  \rho_{S,I}^{(0)}(\vec{n}):\,,
\label{eq:OPEP-full}
\end{align}
where $g_{A}=1.287$ the axial-vector coupling constant (adjusted to account for the Goldberger-Treiman discrepancy)~\cite{Fettes:1998ud}, $f_{\pi}=92.2$~MeV the pion decay constant, and $\rho_{SI}(\vec{n})$ is the spin- and isospin-dependent  density operator,
\begin{align}
\rho^{(d)}_{S,I}(\vec{n}) = & \sum_{i,j,i^{\prime},j^{\prime}=0,1} a^{\dagger}_{i,j}(\vec{n}) \, [\sigma_{S}]_{ii^{\prime}} \, [\sigma_{I}]_{jj^{\prime}} \, a^{\,}_{i^{\prime},j^{\prime}}(\vec{n})
\nonumber \\
& +
s_{\rm L}
\sum_{|\vec{n}-\vec{n}^{\,\prime}|^2 = 1}^d \,
\sum_{i,j,i^{\prime},j^{\prime}=0,1} a^{\dagger}_{i,j}(\vec{n}^{\,\prime}) \, [\sigma_{S}]_{ii^{\prime}} \, [\sigma_{I}]_{jj^{\prime}} \, a^{\,}_{i^{\prime},j^{\prime}}(\vec{n}^{\,\prime}) \,.
\label{eqn:appx--005}
\end{align} with $\vec{\tau}, \vec{\sigma}$ the Pauli-(iso)spin matrices and annihilation (creation) operators $a^{}$ ($a^{\dagger}$). Here, $f_{S',S}$ is
the locally-regulated pion correlation function,
\begin{align}
f_{S',S}(\vec{n}'-\vec{n}) = &\frac{1}{L^3}\sum_{\vec{q}}
\frac{q_{S'}q_{S} \, e^{-i\vec{q}\cdot(\vec{n}'-\vec{n})-(\vec{q}^2+M^2_{\pi})/\Lambda_{\pi}^2}}{\vec{q}^2 + M_{\pi}^2} \,,
\end{align}
where $f^{\pi}$ is a local regulator defined in momentum space,
\begin{align}
f^{\pi}(\vec{n}'-\vec{n})
= &
\frac{1}{L^3}
\sum_{\vec{q}}
e^{-i\vec{q}\cdot(\vec{n}'-\vec{n})-(\vec{q}^2+M^2_{\pi})/\Lambda_{\pi}^2}\,,
\end{align}
with $\vec{q} = \vec{p}- \vec{p}^{\prime}$ the momentum transfer ($\vec{p}$ and $\vec{p}^{\prime}$ are the relative incoming and outgoing momenta). In addition, $C_{\pi}$ is the coupling constant of the OPE counter term given by,
\begin{align}
C_{\pi} = & -
\frac{\Lambda_{\pi} (\Lambda_{\pi}^2-2M_{\pi}^{2}) + 2\sqrt{\pi} M_{\pi}^3\exp(M_{\pi}^2/\Lambda_{\pi}^{2}){\rm erfc}(M_{\pi}/\Lambda_{\pi})}
{3 \Lambda_{\pi}^3}\,,
\end{align}
with $\Lambda_{\pi}=300$~MeV the regulator parameter  and $M_{\pi}=134.98$~MeV the pion mass. Also, $V_{\rm Coulomb}$ is the Coulomb interaction, $V_{\rm 3N}^{\rm Q^3}$ is the 3N potential, $V_{\rm 2N}^{\rm Q^4}$ is the 2N short-range interaction at N3LO,  $W_{\rm 2N}^{\rm Q^4}$ is the 2N Galilean invariance restoration (GIR) interaction at N3LO. For the details of the Coulomb interaction and the two-nucleon (2N) short-range interactions we refer the reader to Ref.~\cite{Li:2018ymw}. The three-nucleon (3N)  interactions at ${\rm Q^3}$ consist of locally smeared contact interactions, one-pion exchange interaction with that the two-nucleon contact terms are smeared locally, two-pion exchange potential~\cite{Friar:1998zt,Epelbaum:2002vt,Epelbaum:2009zsa}, and two additional SU(4) symmetric potentials denoted by $V_{c_{E}}^{(l)}$ and $V_{c_{E}}^{(t)}$. Therefore, the three-nucleon interactions at ${\rm Q^3}$ has the form
\begin{align}
V_{\rm 3N}^{\rm Q^3}
= V_{c_{E}}^{(0)} +
  V_{c_{E}}^{(1)} +
  V_{c_{E}}^{(2)} +   V_{c_{E}}^{(l)} +
  V_{c_{E}}^{(t)} +
  V_{c_{D}}^{(0)} +
  V_{c_{D}}^{(1)} +
  V_{c_{D}}^{(2)}
+ V_{\rm 3N}^{\rm (TPE)}
\,.
\label{eqn:V_NNLO^3N--001}
\end{align}
Here, we first define the two-pion exchange potential, which can be separated into the following three parts,
\begin{align}
V_{\rm 3N}^{\rm (TPE1)}
= & \frac{c_{3}}{f_{\pi}^{2}}\,
\frac{g_{A}^{2}}{4 f_{\pi}^{2}}
\, \sum_{S,S^{\prime},S^{\prime\prime},I}
\sum_{\vec{n},\vec{n}^{\,\prime},\vec{n}^{\,\prime\prime}}	
\, : \, \rho_{S^{\prime},I}^{(0)}(\vec{n}^{\,\prime}) \,
f_{S^{\prime},S}(\vec{n}^{\,\prime}-\vec{n})
f_{S^{\prime\prime},S}(\vec{n}^{\,\prime\prime}-\vec{n})
\rho_{S^{\prime\prime},I}^{(0)}(\vec{n}^{\,\prime\prime}) \,
\rho^{(0)}(\vec{n})
\, : \,
\label{eqn:V_TPE1^3N--001}
\end{align}
\begin{align}
V_{\rm 3N}^{\rm (TPE2)}
= &
-\frac{2c_{1}}{f_{\pi}^{2}}\,
\frac{g_{A}^{2} \, M_{\pi}^{2}}{4 f_{\pi}^{2}}
\, \sum_{S,S^{\prime},I}
\sum_{\vec{n},\vec{n}^{\,\prime},\vec{n}^{\,\prime\prime}}	
\, : \, \rho_{S^{\prime},I}^{(0)}(\vec{n}^{\,\prime}) \,
f_{S^{\prime}}^{\pi\pi}(\vec{n}^{\,\prime}-\vec{n})
f_{S}^{\pi\pi}(\vec{n}^{\,\prime\prime}-\vec{n})
\rho_{S,I}^{(0)}(\vec{n}^{\,\prime\prime}) \,
\rho^{(0)}(\vec{n})
\, : \, \,,
\label{eqn:V_TPE2^3N--001}
\end{align}
\begin{align}
V_{\rm 3N}^{\rm (TPE3)}
=   \frac{c_{4}}{2f_{\pi}^{2}}
&
\left( \frac{g_{A}}{2 f_{\pi}}\right)^{2}
\sum_{S_{1},S_{2},S_{3}}
\sum_{I_{1},I_{2},I_{3}}
\sum_{S^{\prime},S^{\prime\prime}}
\sum_{\vec{n},\vec{n}^{\,\prime},\vec{n}^{\,\prime\prime}}
\varepsilon_{S_1,S_2,S_3}
\varepsilon_{I_1,I_2,I_3}
\nonumber \\
&
\times	
\, : \, \rho_{S^{\prime},I_{1}}^{(0)}(\vec{n}^{\,\prime}) \,
f_{S^{\prime},S_{1}}(\vec{n}^{\,\prime}-\vec{n})
f_{S^{\prime\prime},S_{2}}(\vec{n}^{\,\prime\prime}-\vec{n})
\rho_{S^{\prime\prime},I_{2}}^{(0)}(\vec{n}^{\,\prime\prime}) \,
\rho_{S_{3},I_{3}}^{(0)}(\vec{n})
\, : \, \,,
\label{eqn:V_TPE3^3N--001}
\end{align}
where the locally smeared spin-isospin symmetric density operator is defined as, \begin{align}
\rho^{(d)}(\vec{n}) = \sum_{i,j=0,1} a^{\dagger}_{i,j}(\vec{n}) \, a^{\,}_{i,j}(\vec{n})
+
s_{\rm L}
\sum_{|\vec{n}-\vec{n}^{\,\prime}|^2 = 1}^d \,
\sum_{i,j=0,1} a^{\dagger}_{i,j}(\vec{n}^{\,\prime}) \, a^{\,}_{i,j}(\vec{n}^{\,\prime})
\,,
\label{eqn:appx--001}
\end{align}
and the LECs of two-pion exchange potentials are fixed from pion--nucleon scattering data, $c_{1}=-1.10(3)$, $c_{3}=-5.54(6)$ and $c_{4}=4.17(4)$ all in GeV$^{-1}$~\cite{Hoferichter:2015tha}. We now define the one-pion exchange interaction with the two-nucleon contact terms smeared locally,
\begin{align}
V_{c_{D}}^{(d)}
= -\frac{c_{D}^{(d)} \, g_{A}}{4f_{\pi}^{4} \Lambda_{\chi}}\,  \sum_{\vec{n},S,I}
\sum_{\vec{n}^{\,\prime},S^{\prime}}
\, : \,
\rho^{(0)}_{S^{\prime},I}(\vec{n}^{\,\prime})
f_{S^{\prime},S}(\vec{n}^{\,\prime}-\vec{n})
\rho^{(d)}_{S,I}(\vec{n})
\rho^{(d)}(\vec{n})
\, : \,
\,,
\label{eqn:V_cD-sL-001} \end{align}
and the locally smeared contact interactions as, \begin{align}
V_{c_{E}}^{(d)}= \frac{c_{E}^{(d)}}{6}
\, \sum_{\vec{n},\vec{n}^{\,\prime},\vec{n}^{\,\prime\prime}}
\,
\left[\rho^{(d)}(\vec{n}) \right] ^3\,,
\label{eqn:V_cE-sL-001}      \end{align}
and finally we define two additional SU(4) symmetric potentials denoted by $V_{c_{E}}^{(l)}$ and $V_{c_{E}}^{(t)}$ as,
\begin{align}
V_{c_{E}}^{(l)} = c_{E}^{(l)} \, \sum_{\vec{n},\vec{n}^{\,\prime},\vec{n}^{\,\prime\prime}}
\rho^{(d)}(\vec{n})\,
\rho^{(d)}(\vec{n}^{\,\prime}) \, \rho^{(d)}(\vec{n}^{\,\prime\prime}) \delta_{|\vec{n}-\vec{n}^{\,\prime}|^2,1} \, \,  \delta_{|\vec{n}-\vec{n}^{\,\prime\prime}|^2,1} \, \,  \delta_{|\vec{n}^{\,\prime}-\vec{n}^{\,\prime\prime}|^2,4},  \label{eqn:V_cE-l-001}      \end{align}
\begin{align}
V_{c_{E}}^{(t)} = c_{E}^{(t)}
\, \sum_{\vec{n},\vec{n}^{\,\prime},\vec{n}^{\,\prime\prime}}
\rho^{(d)}(\vec{n})\,
\rho^{(d)}(\vec{n}^{\,\prime}) \, \rho^{(d)}(\vec{n}^{\,\prime\prime}) \delta_{|\vec{n}-\vec{n}^{\,\prime}|^2,2} \, \,  \delta_{|\vec{n}-\vec{n}^{\,\prime\prime}|^2,2} \, \,  \delta_{|\vec{n}^{\,\prime}-\vec{n}^{\,\prime\prime}|^2,2}\,.
\label{eqn:V_cE-t-001}      \end{align}
 It is important to stress that in contrast to the
continuum case, where we just have two LECs, namely $c_E$ and $c_D$, these are smeared here over
neighbouring lattice sites and appear with independent LECs $c_{D,E}^{(0)}, c_{D,E}^{(1)}, c_{D,E}^{(2)},...\,$.
In lattice units, these LECs take the values
\begin{eqnarray}
c_{D}^{(0)} &=& -1.2787~,~~
c_{D}^{(1)} = -2.5665~,~~
c_{D}^{(2)} = -0.2578~,\nonumber\\
c_{E}^{(0)} &=& 3.3724~,~~
c_{E}^{(1)} = 4.9896~,~~
c_{E}^{(2)} = -1.0876~,~~ %\nonumber\\
c_{E}^{(l)} =  -0.4991~,~~
c_{E}^{(t)} = 0.06575~.
\end{eqnarray}
Finally, we note that the pion-nucleon vertices are not smeared, so that we can take the
values of the dimension-two LECs $4c_i$ from Ref.~\cite{Hoferichter:2015tha}.

\end{appendix}

%%%%%%%%%%%%%%%%%%%%%%%%%%%%%%%%%%%%%%%%%%%%%%%
%%%%%%%%%%%%%%%%%%%%%%%%%%%%%%%%%%%%%%%%%%%%%%%

%%%%%%%%%%%%%%%%%%%%%%%%%%%%%%%%%%%%%%%%%%%%%%%
%%%%%%%%%%%%%%%%%%%%%%%%%%%%%%%%%%%%%%%%%%%%%%%

\end{document}